%Paper: 9203217
%From: George Siopsis <siopsis@utkux1.utk.edu>
%Date: Wed, 18 Mar 92 15:13:19 EST

\input phyzzx
\tolerance=10000 \def\half{{\textstyle{1\over2}}}

\def\sixt{{\textstyle{1\over6}}}
\def\pl#1{{\sl Phys.~Lett.~\bf #1B}}
\def\pr#1{{\sl Phys.~Rev.~\bf D#1}}

\def\cpc#1{{\sl Comp.~Phys. Comm.~\bf #1}}
\def\anp#1{{\sl Ann.~Phys.~(NY) \bf #1}}
\def\rmp#1{{\sl Rev.~Mod.~Phys. \bf #1}}
\def\etal{{\it et al.}}
\def\sqstev{$\sqrt s=40$~TeV}
\def\etatwet{$|\eta|<2.8$}
\def\bbar#1{\llap{\phantom#1}^{\scriptscriptstyle(}\bar#1^{
\scriptscriptstyle)}}
\def\bbbarf{\vphantom{\bar f}^{{}^{\rlap{$\scriptscriptstyle($}}}\bar f^{{}^)}
\vphantom{\bar f}}
\REF\cjw{See, e.g., S.~Jadach and B.~F.~L. Ward, \cpc{56} (1990) 351;
\pr{40} (1989) 3852; {\it ibid.} {\bf 38} (1988) 2897, and references therein.}
\REF\ddl{D.~B. DeLaney, \etal, to appear.}
\REF\yfs{D.~R. Yennie, S.~C. Frautschi, and H.~Suura, \anp{13} (1961) 379.}
\REF\sja{S.~Jadach, \etal, \pl{268} (1991) 253; \cpc{} (1992) in press.}
\REF\sjw{S.~Jadach and B.~F.~L. Ward, \pl{} (1992) in press, and to appear.}
\REF\sjb{S.~Jadach, \etal, \pr{42} (1990) 2977.}
\REF\jwb{S.~Jadach and B.~F.~L. Ward, \pr{40} (1989) 3582.}
\REF\eei{See, e.g., E.~Eichten, \etal, \rmp{56} (1991) 579.}
\REF\fep{See, e.g., F.~E. Paige and D.~Protopopescu, in {\sl Snowmass Summer
Study 1986}, ed. R.~Donaldson and J.~Marx (American Physical Society,
New York 1988) p.320; H.-U.~Bengtsson and T.~Sjostrand, \cpc{46} (1987) 43.}
\FIG\fia{The process $e^+(p_1)+e^-(p_2)\to f(q_1)+\bar f(q_2)+\gamma(k_1)+
\dots+\gamma(k_n)$.}
\FIG\fib{The SSC process $q(p_1)+\bbar q(p_2)\to q'(q_1)+
{\bbar q}'(q_2)+\gamma(k_1)+\dots+\gamma(k_n)$, where $q,q'=u,d,s$.}
\FIG\fic{Histogram of the photon multiplicity in $uu\to uu+n\gamma$
for \etatwet: \hfil\break {\it(a)} \sqstev; {\it(b)} $\sqrt s=6.7$~TeV.
Here,  $v_{min}=10^{-6}$-we have shown in Ref.~[\cjw] that the
cross section does not depend on $v_{min}$.}
\FIG\fid{$v$-distribution for $uu\to uu+n\gamma$, where $v=(s-s')/s$
and $s'=(q_1+q_2)^2$ is the squared final $uu$ invariant mass.
Here, \etatwet: {\it(a)} \sqstev; {\it(b)} $\sqrt s=6.7$~TeV.}
\FIG\fie{Total transverse momentum distribution of the photons in
$uu\to uu+n\gamma$ for \etatwet\ in units of $s$: {\it(a)} \sqstev; {\it(b)}
$\sqrt s=6.7$~TeV.}
\TABLE\tbl{Sample output for $uu\to uu+n\gamma$ at \sqstev\ and \etatwet.
The entries in the table are largely explained therein:
XSEC = cross section, WT = event weight, and BORN = Born cross section.}
\Pubnum={UTHEP-92-0101} \date={January 1992} \pubtype={}
\titlepage
\title{Multiple photon effects in fermion-(anti)fermion scattering at SSC
energies${}^\ast$}
\footnote{\ast}{Supported in part by the Texas National Research Laboratory
Commission for the Superconducting Super Collider Laboratory via grant
RCFY9101, and by the Department of Energy Contract No. DOE-AC05-76ER03956.}
\author{D.~B. DeLaney, S.~Jadach,${}^\dagger$ Ch.~Shio, G.~Siopsis, B.~F.~L.
Ward}
\address{Department of Physics and Astronomy\break
The University of Tennessee\break Knoxville, TN 37996--1200\break}
\footnote{\dagger}{Permanent address: Institute of Physics, Jagellonian
University, Cracow, Poland.}
\abstract
We use the theory of Yennie, Frautschi and Suura to realize, via Monte Carlo
methods, the
process $f\,\bbbarf\to f'\,\bbbarf'+n\gamma$ at SSC and LHC energies,
where $f$ and $f'$ are quarks or leptons.
QED infrared divergences are canceled to all orders in perturbation theory.
The resulting Monte Carlo event generator, SSC-YFS2, is used to study the
effects of initial-state photon radiation on these processes in the SSC
environment. Sample Monte Carlo data are presented and discussed.
We find that the respective multiple-photon effects must be taken into account
in discussing precise predictions for SSC physics processes.
\endpage

\chapter{Introduction}

Now that the SSC is under construction, it is extremely important to prepare
for the maximal physics utilization and exploration of the new frontier which
it will probe. In particular, it is of some import to determine the effects of
higher-order radiative corrections on the SSC physics processes of interest so
that optimal discrimination between signal and background can be realized.
In this paper, we explore the first step in the determination of such
corrections by computing them for the basic QED-related effects in
%$ff(\bar{f})\to f'f'(\bar{f}')+n\gamma$ at SSC
$f\,\bbbarf\to f'\,\bbbarf')+n\gamma$ at SSC
energies using the methods which two of us (S.J. and B.F.L.W.) introduced
for the analogous processes $e^+e^-\to f\bar{f}+n\gamma$ for
high precision $Z^0$ physics at SLC and LEP.
Thus, here, we extend the SLC/LEP Monte Carlo event generator YFS2
Fortran in the first paper in Ref.~[\cjw] to the SSC physics environment.
An analogous study of the multiple-gluon radiation in SSC processes such
as $qq\to q'q'+nG$ (where $q,q'$ represent quarks and $G$ a gluon) will
appear elsewhere~[\ddl].

In the SSC environment, multiple-photon and multiple-gluon radiative effects
are expected to be important, in partial analogy with the significance of
multiple-photon radiation in the SLC and LEP environments.
The analogy is partial because the would-be resonance at the SSC, the Higgs,
is actually quite broad compared to the $Z^0$ at SLC and LEP.
Nonetheless, if $\bar{k}_0\approx0.001\sqrt s/2$ is a typical infrared
resolution factor for an SSC detector,
then the probability of an incoming $u$-quark to radiate at the SSC is, e.g.,
$$\eqalign{P(\bar{k}_0\leq k\leq\sqrt{s}/2)&\approx
{2\alpha Q_u^2\over\pi}(\log(\sqrt{s}/6m_u)^2-1)\log(\sqrt{s}/\bar{k}_0)\cr
&=0.44\;\,,\cr}\eqn\pk$$
where \sqstev\ and $m_u\sim5.1$ MeV.
Hence, such radiation and its attendant effect on the respective SSC
event structure must be computed to the standard SLC/LEP precision to
assess its interplay with detector cuts, physics signals and physics
backgrounds
   .
This would then leave only the multiple-gluon radiative effects to be
taken into account to gain a complete view of higher-order radiative
effects to SSC physics processes.
Such gluon radiation will be taken up elsewhere~[\ddl].

Our strategy is to treat the incoming quark radiation via the YFS theory~[\yfs]
so that we realize it on an event-by-event basis using the
methods in Ref.~[\cjw]. The full multiple-photon character of the final state,
including the physical four-momentum vectors of the photons,
is then made available to the users of the attendant new multiple-photon
event generator SSC-YFS2. The implementation of arbitrary detector cuts on the
respective simulated cross sections is straightforward.

A logical next step is to include the effects of the final-state
multiple-photon radiation via the extension of the Monte Carlo event generators
BHLUMI2.0 Fortran~[\cjw,\sja] and YFS3 Fortran~[\yfs] to the SSC environment in
analogy with our extension of YFS2 Fortran in the current work.
We shall discuss these extensions in a future publication~[\ddl].

Our work is organized as follows in the paper: in the next Section, we review
the YFS methods as they are implemented in YFS2 in the first paper in
Ref.~[\cjw], so that this paper is self-contained;
in Section 3, we discuss how we extend YFS2 to SSC processes and
energies to get the Fortran Monte Carlo event generator SSC-YFS2;
in Section 4, we present some sample Monte Carlo data for SSC physics
processes and comment on their implications; finally,
in Section 5, we present our outlook and summary remarks.

\chapter{Review of YFS methods}

In this Section, we review the methods used in Ref.~[\cjw] to realize the
YFS theory via the Monte Carlo event generator YFS2 Fortran for
$e^+e^-\to f\bar{f}+n\gamma$, $f\neq e$, in the $Z^0$ energy regime.
We begin by recalling the key ingredients of these methods.

The YFS Monte Carlo methods in Ref.~[\cjw] take advantage of the expansion of
the total cross section for the process illustrated in Fig.~\fia,
$$e^+(p_1)+e^-(p_2)\longrightarrow f(q_1)+\bar{f}(q_2)+
\gamma(k_1)+\dots+\gamma(k_n)\,\;,\eqn\bsp$$
in terms of the YFS hard-photon residuals~[\yfs] $\bar\beta_n(k_1,\dots,k_n)$,
which are free of all virtual and real infrared divergences to all
orders in the QED coupling constant $\alpha$,
and of the products of the YFS infrared emission factors~[\yfs]
$$\widetilde{S}(k)=-{\alpha\over4\pi^2}\left({p_1\over p_1\cdot
k}-{p_2\over p_2\cdot k}\right)^2+\;\cdots\;\,,\eqn\sti$$
where $\dots$ represents the remaining terms obtained from that shown
by the appropriate substitutions of $\{(e,p_1)\,,\,(-e,p_2)\}$
with $\{(Q_fe,q_1)\,,\,(-Q_fe,q_2)\}$, with due attention to signs
associated with the direction of the flow of charge. The most infrared-singular
contribution to the cross section involves $n$ factors of $\widetilde S$,
as we see from the following expression for the respective cross section:
$$\eqalign{d\sigma^{(n)}&=\left(\widetilde{S}(k_1)\cdots\widetilde{S}(k_n)
\overline\beta_0(p_1,p_2,q_1,q_2)+\dots+\overline\beta_n(k_1,\dots,k_n)\right)
\cr&\times{1\over n!}\delta^4\left(p_1+p_2-q_1-q_2-\sum_{i=1}^n k_i\right)
{d^3q_1\over q_1^0}{d^3q_2\over q_2^0}{d^3k_1\over k_1^0}\cdots
{d^3k_n\over k_n^0}\,e^{2\alpha{\rm Re}B},\cr}\eqn\dsi$$
where $B$ is the YFS virtual infrared function and is given in Refs.~[\cjw]
and~[\yfs]. Basing ourselves on Eq.~\dsi, in YFS2 Fortran we proceed as
follows.

We use the YFS form factor,
$$\eqalign{F_{YFS}(p_1,p_2,\epsilon)&=\exp\left(2\alpha{\rm Re}B+\int{d^3k\over
k^0}\widetilde{S}(p_1,p_2,k)\,(1-\theta(k^0-\epsilon\sqrt s/2))\right)\cr
&=\exp\left({\alpha\over\pi}(2(\ln(s/m_e^2)-1)\ln\epsilon
+\half\ln(s/m_e^2)-1+\pi^2/3\right)\,\;, \cr}\eqn\fyf$$
to compensate for the omission of small-energy photons with $k^0<\epsilon\sqrt
s/2$ for $\epsilon\ll1$ from the phase space in \dsi\ to all orders in
$\alpha$,
as is effected by inserting $\prod_{i=1}^n\theta(2k_i^0/\sqrt s-\epsilon)$ into
\dsi\ and summing over all $d\sigma^{(n)}$. Here we presume we are in the
$e^+e^-$ center-of-mass frame. For YFS2, the hard photon residuals $\overline
\beta_{0,1,2}$ are used, as two of us have explained in Ref.~[\cjw].
This means that, in the YFS2 Monte Carlo itself, some choice must be made
for the reduction of the $n$-photon $+f\bar f$ phase space to the $j$-photon
$+f\bar f$ phase space ($n=0,1,2,\dots$, $j=0,1,2$, $n\geq j$), which is
involved in the definition of the residuals $\overline\beta_i$ ($i=0,1,2$).
We call this choice the reduction procedure ${\cal R}$ and the exact YFS result
$\sum_nd\sigma^{(n)}$ is independent of it, if it is done according to the
rigorous YFS theory. Our choice for ${\cal R}$ is explained in~[\cjw].
Finally, we should emphasize that, for efficient event generation, it is always
desirable to generate a background population of events according to a set of
distributions $d\sigma'^{(n)}$ which embody all of the general features of
Eq.~\dsi, but which remove unnecessary details, and to restore the exact
distributions $d\sigma^{(n)}$ in \dsi\ by rejection methods.
In Refs.~[\cjw] we follow this strategy in constructing YFS2;
in addition, several changes of variables are used to make the background
generation simpler and more efficient from the standpoint of CPU time.
In this way we have realized the YFS theory for $e^+e^-\to f\bar f
+n\gamma$ with $\overline\beta_0$, $\overline\beta_1$, and $\overline\beta_2$,
where we should emphasize that $\overline\beta_2$ has only been included in
the second-order leading-log approximation~[\sjb].

In the next section, we discuss how we extend YFS2 to more general incoming
$f\bar f$ and $ff$ initial and final states, as well as the modifications
needed to make the program applicable in the SSC energy regime.

\chapter{YFS2 at SSC energies}

In this section we describe how one extends the YFS2 Monte Carlo program in
Ref.~[\cjw] to realize particle interaction at SSC energies. Such an extension
involves the introduction of new physics (mainly through a modification of the
Born cross section), numerical problems (due to the very high energies
involved, care is needed for the accuracy of the formulas), and certain
technical problems associated with the Monte Carlo weight rejection method.

We begin by discussing the modifications made to introduce the new
physics at SSC energies. The new program, SSC-YFS2,
computes the cross section for the interaction
$$f(p_1)+\bbbarf(p_2)\longrightarrow f'(q_1)+\bbbarf'(q_2)
+\gamma(k_1)+\dots+\gamma(k_n)\,\;,\eqn\ffn$$ where $f$ is any lepton or quark.
\footnote{\natural}{At present, the program cannot handle third-generation
fermions.} This is still not the most general form of SSC interactions, because
the incoming fermions are of the same type (identical or
particle-antiparticle).
We are currently generalizing the program to include all interactions,
and shall report on the results shortly~[\ddl].

The mass parameters $m_q$ used for the quarks are the Lagrangian quark masses.
We have in mind that the overall momentum transfer in the interactions will be
large compared to the typical momenta inside the proton. In fact, these quark
mass parameters should strictly speaking be running masses $m_q(\mu)$, where
$\mu$ is the scale at which they are being probed. Such a running mass effect
is well-known and is readily incorporated in the program, as the accuracy one
is interested in may dictate. Thus, with this understanding, further explicit
reference to the running mass effect is suppressed.

The interactions realized by YFS2~(Eq.~\bsp) involve only an exchange of
$\gamma$ and $Z^0$ in the $s$-channel. For SSC-YFS2, realizing the more general
interaction~\ffn, $\gamma$, $Z^0$ and $W^\pm$ exchange in the $t$- and
$u$-channels, accordingly, had to be introduced. This was done by generalizing
the Born cross section to include the additional channels. Moreover, in the
case of quark interactions, a gluon exchange was added in all three channels.
The running strong coupling constant
$$\alpha_s(\mu)={12\pi\over(33-2n_f)\ln\{\mu^2/(\Lambda_{n_f}^ %check \{\}
% sizes
{\overline{MS}})^2\}}\,\;,\eqn\sal$$ was used, where $n_f$ is the number of
quark flavors below the energy level $\mu$. In our case, $n_f=6$, and therefore
the QCD parameter $\Lambda_6^{\overline{MS}}$ is used. It can easily be related
to the experimentally measured parameter $\Lambda_4^{\overline{MS}}=238\,MeV$:
$$\Lambda_6^{\overline{MS}}=\Lambda_5^{\overline{MS}}\left({\Lambda_5
^{\overline{MS}}\over m_t}\right)^{\!2/21},\,\;
\Lambda_5^{\overline{MS}}=\Lambda_4^{\overline{MS}}\left({\Lambda_4
^{\overline{MS}}\over m_b}\right)^{\!2/23}.\eqn\lms$$
The masses of the top and bottom quarks were set to $m_b=5$~GeV and
$m_t=250$~GeV, respectively, but the results
are little affected by their precise values.

Certain numerical problems arise at very high energies, because of the very
small value of all ratios $m/\sqrt s$, where $m$ is the mass of any interacting
particle, and \sqstev\ is the energy of the incoming fermions in
their center-of-mass frame.
Certain formulas had to be rewritten so that such small numbers would not be
ignored by the computer when they should not be; if one is not careful, ratios
of the form $0/0$ appear at various places. Working at SSC energies, however,
has the advantage that all terms of order $m^2/s$ or higher can be dropped.
The error is negligible and leads to a considerable simplification of formulas,
and consequently to a reduction in computer time.

Next we discuss the event-generation procedure. To perform the integral for the
total cross section, we first simplify the form of the differential cross
section. Thus the exact cross section $d\sigma$ is replaced by $d\sigma'$,
so that the integral $\int d\sigma'$ can be performed analytically.
The exact cross section is then computed by rejecting events according to
their weights, $$w={d\sigma\over d\sigma'}\,\;.\eqn\wds$$
Apart from simplicity, we require that $d\sigma'$ lead to an efficient
generation of events. In YFS2, $d\sigma'$ was chosen to be a constant.
In the present case, this is no longer possible, because of the presence of the
$t$-channel. The cross section has a singularity at $t\equiv(p_1-q_1)^2=0$ of
the form $1/t^2$. To account for the singularity, an angle cutoff
$\theta_0=100$~mrad is introduced. This is in accord with current detector
capabilities, and can be changed at will. A crude cross section $d\sigma'$ of
the form $$d\sigma'=A+{B\over t^2}\;,\eqn\scr$$ was chosen,
where the constants $A$ and $B$ depend on the interaction.
\footnote{\ddagger}{A fictitious photon mass cutoff was also tested,
but it turned out to lead to a large weight rejection rate.}
When a $u$-channel also contributes, a similar term of the form $1/u^2$ must
be added to account for the singularity at $u\equiv(p_1-q_2)^2=0$.

Finally, we comment on the choice of the reduction procedure which is needed
for the definition of the arguments of the YFS residuals $\overline\beta_i$
($i=0,1,2$), as explained in Section~2. The reduction procedure is more
delicate in the presence of the $t$-channel, due to the singularity at $t=0$.
One has to make sure that the weights~\wds\ do not become uncontrollably large.
This is managed by making $t$ as large as possible after the reduction.
It is not always possible to increase the reduced $t$ so that the weight~\wds\
remains below the maximum weight. The object of this exercise is to minimize
the error originating from the tail of the distribution of weights above the
maximum weight (which is set to 3, but can be changed if so desired).
This is accomplished by a somewhat involved reduction
procedure, which is an adaptation of the similar procedure in
BHLUMI1.xx~[\jwb].

This concludes our discussion of the modifications in the YFS2 program
necessary in order to realize interactions at SSC energies.
Next, we present some of our results.

\chapter{Multiple-photon effects at SSC energies}

In this Section we present some results on the effects of multiple-photon
initial-state radiation on the incoming $qq$ and $q\bar q$ ``beams'' at
SSC energies using our YFS Monte Carlo event generator SSC-YFS2 Fortran.
Our objective is to determine the size of these effects with an eye toward
their incorporation into SSC physics event generators.
This latter step will be taken up elsewhere~[\ddl].

We consider, as illustrated in Fig.~\fib\ (the kinematics is summarized
in the figure),
$$q(p_1)+\bbar q(p_2)\longrightarrow q'(q_1)+{\bbar q}'(q_2)
+\gamma(k_1)+\dots+\gamma(k_n)\,\;,\eqn\qqn$$ at \sqstev\ for $q,q'=u,d,s$.
For definiteness, we will illustrate our results with $q=u,d$, where we use
$m_u=5.1\times10^{-6}$~TeV, $m_d=8.9\times10^{-6}$~TeV, and view \sqstev\ as
our worst-case scenario. The more typical~[\eei] value $\sqrt
s\approx\sixt40$~TeV $\approx6.7$~TeV is also presented here for completeness.
For these respective input scenarios, we shall discuss the following
distributions: the number of photons per event, the value of $v=(s-s')/s$,
where $s'=(q_1+q_2)^2$, and the squared transverse momentum of the outgoing
$n\gamma$ state.  These distributions give us a view of the effect of this
multiple-photon radiation on the incoming quarks and (anti)quarks in the SSC
environment, where one is really interested in $p\bar p\to H+X$,
where $H$ is the Standard Model Higgs particle.

Considering first the number of photons per event, we have the results in
Fig.~\fic. There, we show that for the $uu$ incoming beams,
the mean number $\VEV{n_\gamma}$ of radiated photons is $0.85\pm0.92$
(it is similar for $\sqrt s=6.7$~TeV). This should be compared to the $dd$
incoming state, where $\VEV{n_\gamma}$ is $0.21\pm0.45$.
For reference, we recall~[\cjw] that at LEP/SLC energies,
the corresponding value of $\VEV{n_\gamma}$ is, for the incoming
$e^+e^-$ state, $\sim1.5\pm1.0$. Hence, we see here one immediate effect of the
high energy of the SSC incoming beams: the initial $uu$-type state will radiate
a significant number of real photons, with a consequent change in the observed
final-state character. In particular, the issue of how much energy is lost to
photon radiation is of immediate interest,
for this energy is unavailable for Higgs production by $uu$ (or $dd$)
and, further, it may fake a signal of $H\to\gamma\gamma$ if we are unlucky.
Accordingly, we now look at the predicted distribution of
$$v\equiv(s-s')/s\,\;,\eqn\vssp$$ where $s'=(q_1+q_2)^2$ is the squared
invariant final fermion pair mass. If only one photon is radiated,
$v$ is just the energy of this photon in the center-of-mass system of the
incoming beams (in units of the incoming beam energy).

What we find for $v$ is shown in Fig.~\fid\ for the $uu\to uu+n\gamma$ case
(the $dd\to dd+n\gamma$ case is similar). We see the expected shape of $v$ from
Ref.~[\cjw], and its average value is $\VEV{v}=0.05\pm0.09$.
% ($0.05\pm0.09$) for $\sqrt s=40$ ($6.7$)~TeV.
Hence, $\sim10\%$ of the incoming energy is radiated into photons;
this energy is not available for Higgs production and hence it is crucial to
fold our radiation into the currently available SSC Higgs production Monte
Carlo event generators~[\fep] and to complete the development of our own
YFS multiple-photon (-gluon) Higgs production Monte Carlo event generator,
which is under development and will appear elsewhere~[\ddl].

Given that we know we have, in the SSC environment, significant multiple-photon
radiation effects, the question of immediate interest is how often the
transverse momenta of two photons are large enough that they could fake a
$H\to\gamma\gamma$ signal. We will answer this very important question in
detail in the not-too-distant future when our complete Higgs production YFS
Monte Carlo event generators are available~[\ddl]. However, here we can begin
to study this question by looking into the transverse momentum distribution of
our YFS multiple-photon radiation in, e.g., $uu\to uu+n\gamma$.
This is shown in Fig.~\fie, where we plot the         total transverse
momentum distribution of the respective YFS multiple-photon radiation.
What we find is that
, for $\sqrt s = 40$~TeV,
the average value of this total   transverse momentum is
(in the incoming $uu$ center-of-mass system)
$$\VEV{p_{\perp,tot}}\equiv\VEV{|\sum_{i=1}^n\vec{k_{i\perp}}|}
=(0.0184\pm0.0129)\sqrt s\,\;,\eqn\pper$$
where $k_i$ ($i=1,\dots,n$) are the four-momenta of the $n$ photons.
  (For $\sqrt s=6.7$~TeV, this average is $(0.0186\pm0.0136)\sqrt s$.)
Hence, for the SDC acceptance cut of $\half|\ln\tan(\theta/2)|\equiv
|\eta|<2.8$, or $\theta_i>122$~mrad, this means that there may be some possible
background to $H\to\gamma\gamma$ for, e.g., $m_H\approx150$~GeV.
Such effects will be discussed in detail elsewhere~[\ddl].

Finally, concerning the overall normalization, we find that the Born cross
section is corrected according to the results in Table~\tbl. This shows clearly
that the higher-order effects change the normalization by approximately~$5\%$.
This sets the level at which precise simulations of SSC physics must take the
higher-order effects from multiple photons into account.

\chapter{Conclusions}

We conclude that our initial study of YFS multiple-photon radiation in the SSC
physics environment shows that any Monte Carlo event generator which hopes to
achieve an accuracy of order~$10\%$ in the SSC physics simulations must treat
the respective effects in a complete way.
In this paper, we have computed these effects for incoming quark-(anti)quark
states at SSC energies using the Monte Carlo event generator SSC-YFS2 Fortran
based on our original YFS2 Monte Carlo in Ref.~[\cjw].

Specifically, using our SSC-YFS2 Monte Carlo event generator for $q\bbar q\to
q'{\bbar q}'+n\gamma$, at \sqstev, we find that for an initial $uu$ state,
the mean number of radiated photons is $0.85\pm0.92$, so that the
multiple-photon character of the events must be taken into account in detailed
detector simulation and physics analysis studies. Further, the mean value of
$v=(s-s')/s$ is $0.05\pm0.09$ and the average total squared transverse momentum
$\VEV{k_{\perp,tot}^2}$ is $0.025\pm0.002$~s.
Hence, the impact of these event characteristics on Higgs production in general
and on the $H\to\gamma\gamma$ scenario in particular must be assessed in
detail.
Such assessment will appear elsewhere~[\ddl].

In conclusion, we can say that the initial platform for precision SSC
electroweak physics simulations on an event-by-event basis using our YFS
Monte Carlo approach~[\cjw] has now been established. We look forward with
excitement to its complete development for all such electroweak phenomena and
to its extension to the SSC QCD processes as well~[\ddl].

\vskip5mm

\noindent{\bf Acknowledgements}

The authors have benefited from the kind hospitality of Prof.~F.~Gilman of
the SSC Laboratory, where part of this work was done. The authors also thank
Profs.~G.~Feldman, J.~Dorfan and M.~Breidenbach of Harvard and SLAC, and
Prof.~J.~Ellis of CERN for the kind hospitality of the Mark~II and SLD
Collaborations and the CERN Theory Division, where the basic ideas in this
paper were conceived in the context of LEP/SLC $Z^0$ physics.

\vfill\eject\refout\endpage\figout\tabout\endpage\bye